# Association Free Energies of Metal Cations with Mesylate and Acetate in Brine Calculated via Molecular Dynamics Simulation

*Sung Hyun Park[1,2], Rikkert J. Nap[1,2], and Igal Szleifer[1,2,3]\**

Department of Biomedical Engineering[1], Department of Chemistry[2], and Chemistry of Life Processes Institute[3],

Northwestern University, Evanston, IL 60208, U.S.A.

ABSTRACT

Specific interactions between metal cations and negatively charged functional groups in polyelectrolytes can have a significant impact on the macroscopic behavior of polymer systems. An example is the dispersion stability and adsorption properties of polyelectrolyte-coated nanoparticles under high salinity conditions in brine. Here we report results from fully-atomistic molecular dynamics simulations and present the calculated binding free energies of the associations between metal cations and anionic functional groups under brine condition.

Specifically the ion pair formations of $Na^+$ and $Ca^{2+}$ metal cations with sulfonate and carboxylate functional groups were considered. In our simulations, sulfonate and carboxylate functional groups in polyelectrolytes were modeled by mesylate and acetate anions, respectively. The simulations show that the binding strengths of $Na^+$ and $Ca^{2+}$ with sulfonate group are relatively weak with correlation times for the contact ion pairs being around 300-400 ps. In the case of acetate, the binding strength with $Na^+$ was found to be comparable to the case of sulfonate group, while the binding of acetate with $Ca^{2+}$ was found to be much stronger with a correlation time of the order of hundreds of nanoseconds. The binding free energies of $Na^+$ and $Ca^{2+}$ with sulfonate as well as that of $Na^+$ with acetate have been calculated directly from the radial distribution functions. For the case of binding between $Ca^{2+}$ and acetate constrained MD simulations with umbrella sampling were carried out to improve the sampling of the phase space and the potential of mean force was obtained by the weighted histogram analysis method. The binding free energy of a 2:1 binding event between one $Ca^{2+}$ ion and two acetates was also calculated. Finally the binding free energies between $Ca^{2+}$ and acetate were compared for different force field parameters and water models, and also with experimental and calculated values reported in the literature.

INTRODUCTION

An ion pair can form when oppositely charged free ions or charge centers coexist in a solution. A simple example of ion pair formation is the association of free cations and free anions species in an electrolyte solution. Ion pairing, a process of forming an ion pair, is sometimes called ion condensation when the process involves collective associations between multiple charge centers in large molecules such as polyelectrolytes, or proteins, and free counterions in the bulk phase. Ion pairs can be found in a wide variety of systems. In biology the specific effects of salts on protein stability and solubility, known as the Hofmeister effects, are a prime example where advancing the knowledge on the delicate properties of the ion pairings between a protein and different counterions are likely a key element to fully understand the underlying mechanism of the phenomenon.[1] In other charged soft matter systems, such as a surface modified by tethered polyelectrolytes, the condensation or release of the counterions on the charge centers is one of the main processes for structural changes as well as charge regulation that allow the system to determine the equilibrium state in a dynamical fashion upon hanges in the environment.[2]

The structure of an ion pair is determined by a delicate balance of the interactions among ion-ion, solvent-ion, and solvent-solvent pairs in the system, which is often described by three distinct geometries proposed by Eigen and Tamm[3]: a contact ion pair (CIP) where two associating ions are in direct contact with each other, a solvent-shared ion pair (SIP) where two ions are separated by a single layer of water molecules, and a solvent-separated ion pair (SSP) where each of the two ions maintains an intact first hydration layer. These structures have been extensively studied in the past decade through various experimental techniques including neutron scattering,[4] X-ray diffraction,[4a, 5] direct force measurement,[6] and dielectric relaxation spectroscopy.[7] Computational studies of ion pairs, mainly using molecular dynamics (MD) or Monte Carlo (MC) simulations, have also been actively pursued during the same period and have advanced the understanding of ion pairs by providing detailed information regarding structural and thermodynamic characteristics among others.[8]

Superparamagnetic iron oxide ($Fe_3O_4$) nanoparticles (SPIONs), used successfully in biomedical applications as contrast agents in magnetic resonance imaging (MRI), are relatively simple to prepare and have excellent nuclear magnetic resonance (NMR) T2 contrast at low concentrations. There is an interest in extending their applications in enhanced oil recovery (EOR) at oil field. To this end, the nanoparticles need to be stabilized and one way of doing that is by coating polyelectrolytes to their surfaces. The polymer-tethered nanoparticles can be used for a variety of potential oil-field applications such as enhanced imaging and detection of oil in the oil reservoir, as a cargo carrier of tailored surfactants, and as temperature/pressure/viscosity sensors in the oil well. An immediate challenge is to find coating conditions for nanoparticles which can survive the harsh environment of an oil reservoir where the high temperature and pressure as well as the extremely high salinity and complex geochemistry could negatively impact the properties of the nanoparticles in intended applications. The research strategy is centered on coating nanoparticle surfaces with proper polymers in order to achieve desired properties in the reservoir conditions. The most promising polymers in this endeavor are charged polymers based on sulfonates, such as PAMPS (poly(acrylic acid-2-acrylamido-2methylpropane sulfonate)). Nanoparticles coated with PAMPS have shown, in controlled laboratory conditions, to remain well dispersed in high-salinity brine solution. Furthermore, the polymeric coating displayed desirable behavior in preventing nanoparticle adsorption to rock surfaces often found in oil reservoirs.[9] However, attaching PAMPS to nanoparticles often employs acrylic acid (AA) binding mechanisms which could have a detrimental effect on the nanoparticle dispersion stability due to a strong interaction between carboxylate groups on acrylic acids and multivalent ions such as $Ca^{2+}$ that are prevalent in the oil reservoirs.

In order to address some of the questions surrounding the ion pairs that are relevant for the (PAMPS-AA)-coated nanoparticles inside brine as commonly found in oil wells, we carried out a series of molecular dynamics (MD) simulations of $Na^+$ or $Ca^{2+}$ metal cations and negatively charged sulfonate or carboxylate groups to obtain insight in the formation and strength of ion-pairs. Some of the results we present here are being used by a molecular theory[10] to make predictions on the macroscopic properties of the (PAMPS-AA)-coated nanoparticle inside brine, which we plan to present in a separate publication in the near future. The accuracy of the theoretical predictions critically depends on how the theory properly describes important interactions such as the ion pairing of $Na^+$ or $Ca^{2+}$ with charged functional groups in polyelectrolytes such as PAMPS-AA copolymer under high salinity environment in brine. In particular, a critical piece of information needed for the theory to accurately describe the ion pairing is the free energy associated with the binding process. Below we present the results from our MD simulations including the structures and calculated binding free energies of the relevant metal-anion ion pairs in the (PAMPS-AA)-coated nanoparticles. In particular, the binding free energy for a potential 2:1 bridging event between a $Ca^{2+}$ cation and two acetate anions, which is very important but was never studied quantitatively by using MD simulation before to our knowledge, was also calculated in this study. The results for the binding free energies are compared with experimental data available in the literature, which allows us to examine whether MD simulation can be a viable option to obtain binding free energy of ion pair accurately.

SIMULATION METHODS

1. Regular Molecular Dynamics Simulations

In order to simulate the ion pairing events of $Na^+$ or $Ca^{2+}$ metal cations with sulfonate functional groups in brine, all-atom MD simulations have been carried out for a single anionic mesylate molecule ($CH_3SO_3^-$), a deprotonated form of methanesulfonic acid, as a model system under different salt concentrations which combine to brine conditions. Specifically, a negatively charged mesylate molecule with a fully atomistic resolution was initially placed at the center of a 5x5x5 $nm^3$ cubic box, which was then filled with ~3800 explicit water molecules together with either 1.5 M NaCl or 0.18 M $CaCl_2$. Each simulation system was prepared to be charge neutral by adjusting the number of $Cl^-$ ions as necessary. In a separate simulation, we also prepared a mesylate in the presence of a salt mixture of 1.5 M NaCl and 0.18 M $CaCl_2$ to test if the binding energy of the ion pair between one metal cation and sulfonate is affected by the co-presence of another metal cation in the medium. All bonding and nonbonding interactions of the mesylate molecule as well as of $Ca^{2+}$ and $Cl^-$ ions were modeled using the OPLS-AA force field,[11] while the nonbonding interaction parameters of $Na^+$ were taken from the Aqvist parameters.[12] The structure and interactions of the explicit water molecules were described by the SPC water model.[13]

Similar protocols were taken to prepare the systems for the MD simulations of acetate with $Na^+$ and $Ca^{2+}$ metal cations. Again, a negatively charged acetate ($CH_3COO^-$) molecule was described by the OPLS-AA force field with a fully atomic resolution. Each simulation system was set up by initially placing an acetate molecule at the center of a 5x5x5 $nm^3$ simulation box, which was then filled with explicit water molecules together with either 1.5 M NaCl or 0.18M

CaCl$_2$. In the case of the Ca$^{2+}$-acetate systems, we also tested AMBER force field[14] to describe acetate paired with TIP3P[15] as well as with TIP4P[16] water model to compare the effects of different force fields.

In all the regular MD simulations for the sulfonate-metal cation systems, each of the prepared systems was first energetically stabilized by the steepest descent algorithm, which was followed by a short 1-ns equilibration and a 50-ns simulation for production. In the case of the acetate-metal cation systems, the time scales of the binding-unbinding events between the metal cations and acetate have been found to be much longer than for the sulfonate cases. Therefore the regular MD simulations for the acetate-metal cation systems have been run for 1 microsecond for production in order to observe some of the binding-unbinding events of the corresponding ion pairs. All simulations were run on GROMACS package, version 4.6.1.[17] The simulations were run both at a room temperature of 300K as well as at a typical brine temperature of 400K to examine the potential effect of temperature on the binding energy. After energy minimization, the atom positions of the mesylate and the acetate anions were constrained to remain fixed over the course of the production runs for computational efficiency. Particle Mesh Ewald electrostatics[18] was used with a distance cut-off of 1.2 nm, and periodic boundary conditions were applied in all three directions. The Nose-Hoover thermostat[19] was used for temperature control, and the pressure was maintained at 1 atm via the Parrinello-Rahman barostat.[20] The system trajectories were recorded every 1 ps in the simulations for post-simulation analyses. A typical snapshot of the simulated system of a mesylate anion with 1.5 M NaCl and 0.18 M CaCl$_2$ is shown in figure 1.

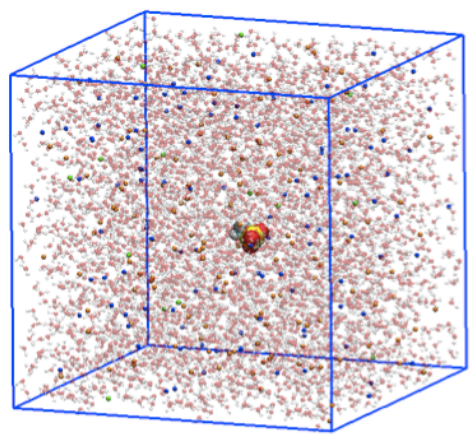

Figure 1. A snapshot of mesylate, at the center of the box, dissolved in water with 1.5 M NaCl and 0.18 M CaCl$_2$ from a regular MD simulation. The temperature is 300K, the pressure is 1 atm, and the initial box size is 5x5x5 nm$^3$. Shown together with mesylate are Na$^+$(blue dots), Ca$^{2+}$(green dots), Cl$^-$(red dots) ions among ~3800 water molecules.

2. Constrained MD simulations with umbrella sampling

The constrained MD simulation with umbrella sampling method[21] was used to calculate the potential of mean force (PMF) of acetate against Na$^+$ and Ca$^{2+}$ metal cations. In our umbrella sampling simulations for 1:1 metal-acetate association events, an acetate molecule and a metal cation, Na$^+$ or Ca$^{2+}$, was solvated with explicit water molecules in a 5x5x5 nm$^3$ cubic box. The distance between the carbon atom in the carboxylate group of the acetate anion and the metal cation was chosen as the reaction coordinate. To perform the umbrella sampling, a series of 31 initial configurations were generated along the reaction coordinate with a separation distance between the acetate and the metal cation increasing by increments of 0.5 Å between two adjacent configurations. The separation distance between the acetate and the metal ion in each of the 31 prepared configurations was then constrained by a harmonic potential in the series of constrained MD simulations. The force constant of the harmonic potential was chosen at k = 20000 kJ/(mol

nm$^2$) following the work by Kaster and Thiel[22], in order to ensure adequate overlaps between adjacent configurations and to achieve proper sampling of configurations at the barrier regions where sampling of the phase space is poor in regular MD simulation. We also carried out umbrella sampling simulations for 1:1 binding events between a 1:1 complex of a Ca$^{2+}$-acetate pair and another acetate anion by using a similar procedure as a way to obtain the free energy of a 2:1 bridge complex formation between two acetate anions and one Ca$^{2+}$ ion. In this case, a 1:1 contact ion pair between an acetate anion and a Ca$^{2+}$ cation was initially placed at the center of the simulation box and another acetate anion was added to the system at incrementally changing distances along a reaction coordinate. After the completion of the simulations, the acetate-Ca$^{2+}$ 1:1 complex was confirmed to remain intact during the entire umbrella sampling simulations between the 1:1 complex and another acetate anion.

The umbrella sampling simulation was run for 1 ns for each constrained distance at a constant pressure of 1 atm and at temperatures of 300K and 400K. After the simulations were completed, the bias of the harmonic constraint was removed properly by the weighted histogram analysis method[23] implemented in the g_wham analysis program[24] in the GROMACS package, and the PMF was obtained.

RESULTS AND DISCUSSION

1. Cation-sulfonate ion pairs

The formation of the ion pairs in the ion condensation process between the metal cations and the sulfonate group from the regular MD simulation can be confirmed by the radial distribution function of the metal cations around the sulfonate group in the mesylate molecule. The radial distribution functions, g(r), of the Na$^+$ and Ca$^{2+}$ ions as a function of radial distance from the sulfonate oxygen atoms were calculated in the following fashion. From the regular MD simulations, the proximal distances between all the metal cations and the three oxygen atoms in the sulfonate group were calculated and a distance histogram for each metal cation was created from them. Each bin in the distance histogram was then divided by the corresponding shell volume of the sulfonate oxygen atoms to calculate a number density of the metal cations around the sulfonate oxygens. Finally the density at a long distance was set to 1 to obtain normalized radial distribution functions. A similar method was used by Van der Vegt *et al.*[8a] for ion pairs between acetate and alkali metals, where they used an analytical geometric expression to obtain proper shell volumes for normalization of the radial distribution function. In our case, the VOIDOO software package[25] was used to calculate the proper shell volumes of the sulfonate oxygen atoms numerically from their Van der Waals radii (VdW). The obtained radial distribution function displays a detailed distribution of the ion pairs with different structures along the radial distance between the negatively charged sulfonate group and the metal cations. The results are shown in figure 2.

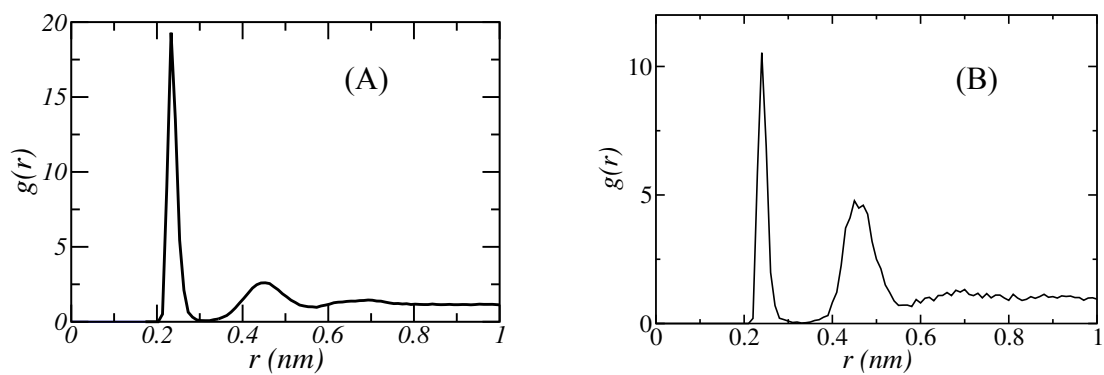

Figure 2. The radial distribution functions, g(r), of (A) $Na^+$ and (B) $Ca^{2+}$ along the radial distance from sulfonate oxygen atoms, obtained from the regular MD simulations.

In the figure, for both $Na^+$ and $Ca^{2+}$ cases, the largest peak at the distance of around $r = 0.22$ nm represent the contact ion pair (CIP) in which two oppositely charged species are in direct contact with each other. The second largest peak near $r = 0.46$ nm correspond to the solvent shared ion pairs (SIP) where two oppositely charged species are separated by a single shared layer of water molecules. Finally the smaller peak at around $r = 0.7$ nm are the solvent separated ions pairs (SSP), where each charged species in an ion pair is surrounded by a single intact hydration layer. The radial distribution functions reveal a small but interesting difference in the relative preference of the ion pair types between the two metal ions. For example, despite having a smaller charge, $Na^+$ ions form slightly more CIPs with the sulfonate group than $Ca^{2+}$ ions do, while the $Ca^{2+}$ ions form relatively more SIPs compared with $Na^+$. This suggests that the first hydration shell of $Ca^{2+}$ may be tighter and more stable than the $Na^+$ counterpart, making $Ca^{2+}$ less likely than $Na^+$ to lose the hydration shell entirely and form a CIP with sulfonate anion. The VdW radius of the $Ca^{2+}$ is indeed smaller than that of Aqvist $Na^+$ in the OPLS-AA force field, so it is reasonable that $Ca^{2+}$ with a smaller size and a higher charge possesses a tighter and more stable hydration shell. The radial distribution functions also offer a convenient and rigorous way to distinguish among CIP, SIP, and SSP based on simple distance criteria. Specifically, in the case of the sulfonate-$Na^+$ system, the sulfonate group and a $Na^+$ cation form a CIP when the distance between the two species is equal to or less than 0.31 nm, the distance where the radial distribution function is minimum between two peaks representing the CIP and SIP state in figure 2(A). When the distance between the sulfonate and a $Na^+$ ion is 0.31 and 0.56 nm, the two species are in the SIP state for that particular moment. When the distance is between 0.56 and 0.81 nm, they form an SSP. When the distance is larger than 0.81 nm, they do not form an ion pair. We will use these distance criteria later to define the states of all the ion pairs from our simulations when we calculate the residence time correlation functions of the ion pairs.

The snapshots from the simulation of a sulfonate anion with both Na$^+$ and Ca$^{2+}$ metal cations suggest that the local structure and stoichiometry of the ion pairs can be very complex. As shown in figure 3, more than one metal cation can simultaneously be associated with a single sulfonate group to form multiple ion pairs in the mixture of two different kinds of metal cations and a sulfonate anion. These complex structures are presumably stabilized by a complex interplay among sulfonate and metal cations as well as local water molecules and co-solvated Cl$^-$ counterions (invisible in the snapshots in figure 3).

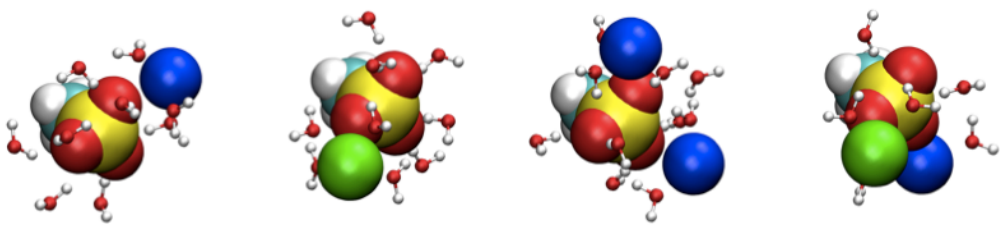

Figure 3. Snapshots of metal-sulfonate ion pairs from MD simulation. (red=oxygen, blue = Na$^+$, green = Ca$^{2+}$, yellow = sulfur, cyan=carbon, white=hydrogen)

The temporal persistency or duration of ion pairs can be used as an indicator of the binding strength of the ion pairs. One way to measure the temporal persistency of the ion pair is to compute the residence time correlation function, $C(t)$, defined as follows:

$$C(t) = \frac{<R(t)R(0)>}{<R^2(0)>}$$

Here, $R(t)$ is a binary operator which is defined as $R(t) = 1$ when an ion-pair is formed, and $R(t) = 0$ when there is no ion-pair formed between a given cation and a given anion at a particular time $t$. The bracket represents an ensemble average, or equivalently, a time average over the course of the simulation. Whether an ion-pair is formed or not between a given metal cation and the sulfonate anion at a particular time point is determined based on the distance criteria for the different ion pair states shown in figure 2. The values for *R(t)* operator can be determined in the following way. First, the distance between a given metal cation and the sulfonate anion at a given time is calculated from the simulation. Next, the calculated distance is compared with the distance ranges for different ion-pair states as shown in the radial distribution functions in figure 2. If the calculated distance falls into one of the three distance ranges for CIP, SIP, or SSP states as defined in figure 2, the value of the binary operator *R(t)* for that particular cation at the given time becomes 1 for the corresponding ion pair state and 0 for the other two ion pair states as well as for the state of no ion pair formation. The residence time correlation function for each ion pair state can be calculated by taking the average values of the product, *R(t)R(0)*, for all cation-sulfonate pairs for all possible time intervals and then by normalizing properly. The residence time correlation functions for the three different states of the ion pairs, i.e. CIP, SIP, and SSP, between a sulfonate anion and the metal cations are shown in figure 4.

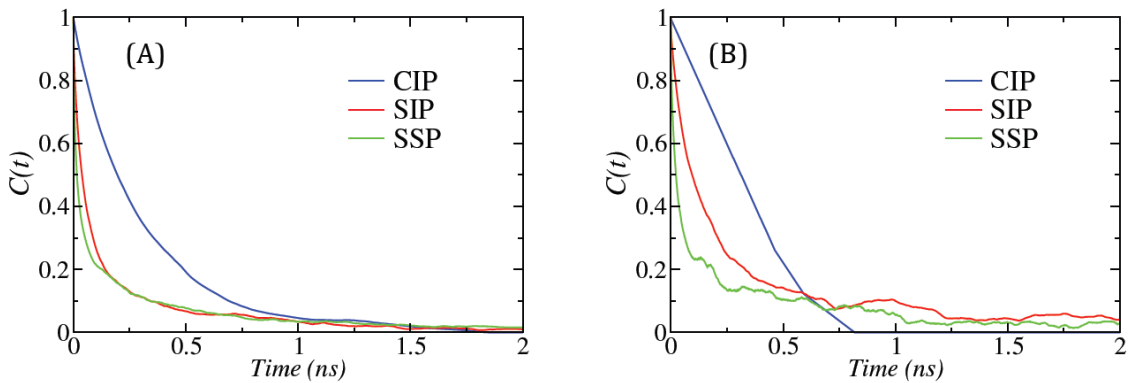

Figure 4. Residence time correlation functions between sulfonate and (A) $Na^+$, or (B) $Ca^{2+}$.

The figure shows that the correlation functions of all sulfonate ion pairs with both $Na^+$ and $Ca^{2+}$ cations decay within the order of hundreds of picoseconds or less. In both $Na^+$ and $Ca^{2+}$ cases, the correlation functions of the CIPs decay slower than those for the SIP, which then decays even slower than SSP, reflecting the overall hierarchy of lifetimes or binding persistency among different ion pair states. As expected, the CIP state with direct contact between oppositely charged ions is the longest-lived, while the SSP state with an intact first hydration layer for each ion the shortest-lived. The correlation times for different ion pairs, defined as the time needed for the correlation function, $C(t)$, to decay to a value of $1/e$, are summarized in Table 1. The results show that the ion pairs between sulfonate and $Na^+$ or $Ca^{2+}$ metal cations from the regular MD simulations persist, on average, for the time durations ranging from around 30 ps to as long as about 400 ps, depending on the specific state of the ion pairs. The sulfonate ion pairs with $Ca^{2+}$ persist slightly longer in all three ion-pair states than those with $Na^+$, however the differences are not very large.

**Table 1.** The residence correlation times of the metal cation-sulfonate ion pairs from the regular MD simulations (unit: ps).

|            | CIP | SIP | SSP |
|------------|-----|-----|-----|
| $Na^+$     | 290 | 65  | 33  |
| $Ca^{2+}$  | 395 | 165 | 44  |

The dynamic characteristics of the ion pairing processes can also be directly captured from the regular MD simulations by extracting the time trajectories of the distances, $L(t)$, between the sulfonate group and the metal cations, as shown in figure 5. Each solid line in the plots represents a distance trajectory of a distinct individual metal cation over the simulation time. The blue and red dashed horizontal lines mark the peak positions of the CIP and SIP ion pair states, respectively, from the radial distribution functions in figure 2. Since the simulations were performed in brine conditions where the $Na^+$ concentration is much higher than $Ca^{2+}$ concentration (1.5M $Na^+$ vs. 0.18 M $Ca^{2+}$), the distance trajectories for $Na^+$ in figure 4a appears more densely populated with more trajectory lines compared with the case for $Ca^{2+}$ in figure 4b. The lines in the figure show that both $Na^+$ and $Ca^{2+}$ ions make rapid and dynamic transitions between different ion pair states when associated with the sulfonate anion over the course of the simulations. Multiple exchanges between the CIP and SIP states are clearly visible in the trajectories for both cation species, while the SSP states around $L(t) = 7$ nm are rather vague to identify in both cases. A closer look at the time trajectories of the cation-sulfonate distances reveals that the scale of the lifetime of the CIPs ranges in the order of 100 ps to 1 ns for both $Na^+$ and $Ca^{2+}$. This is consistent with both the residence time correlation functions, $C(t)$, in figure 4 and the correlation times shown in Table 1. In the case of $Na^+$ in figure 4a, we also observe that more than one $Na^+$ ion frequently form CIP or SIP ion pairs with a single sulfonate group simultaneously at a given time. This observation suggests that the ion pair structure may be much more complex and diverse than just a simple case of a 1:1 ion pair. It also suggests that a complex balance of interactions in such a transient structure probably involves $Cl^-$ counterions and some local water molecules in order to achieve a certain level of stability. Such an association of multiple cations with a single sulfonate is not observed for the $Ca^{2+}$, presumably due to the higher formal charge of $Ca^{2+}$ ions, which should strongly disfavor the formation of ion clusters with multiple $Ca^{2+}$ ions associated with a single sulfonate. The lower concentration of the $Ca^{2+}$ in the simulation should also make such events less likely to occur by lowering the chance for multiple $Ca^{2+}$ ions to assemble simulteneoulsy with a single sulfonate as well as multiple counterions and water molecules.

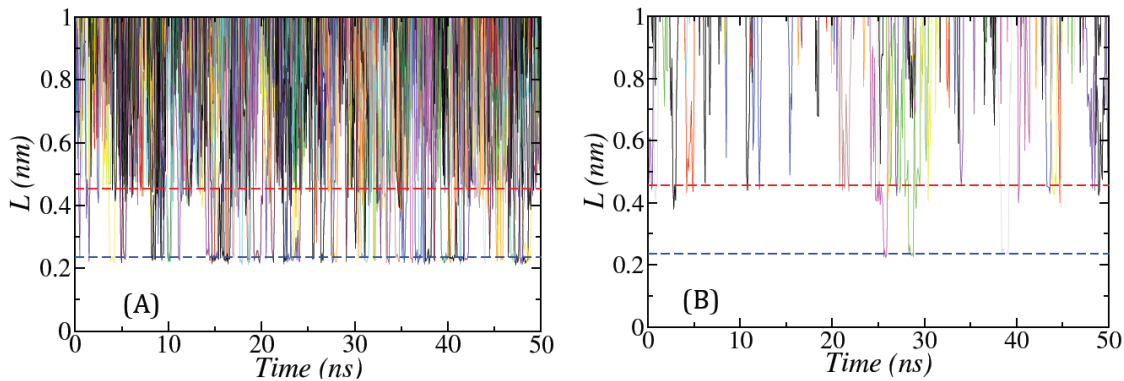

Figure 5. The temporal profiles of the metal cation-sulfonate distances, $L(t)$, for (A) Na$^+$, and (B) Ca$^{2+}$ from the regular MD simulations.

The lifetimes for the cation-sulfonate ion pairs, in the order of 100 ps, are much shorter than the total simulation time length of the MD runs of 50 ns. This means that the sampling of the phase space of ion pair formation and breaking events in the simulations should have decent statistics, and therefore the free energy of the corresponding process of ion pair formation, $\Delta G(r)$, can be directly calculated from the radial distribution function, $g(r)$, by using the following reversible work theorem from statistical mechanics:[26]

$$\Delta G(r) = -k_B T \ln g(r) \qquad (1)$$

Accordingly, we obtained the binding free energy profiles of the Na$^+$-sulfonate and Ca$^{2+}$-sulfonate ion pairs directly from the radial distribution functions, which are shown in the figure 6.

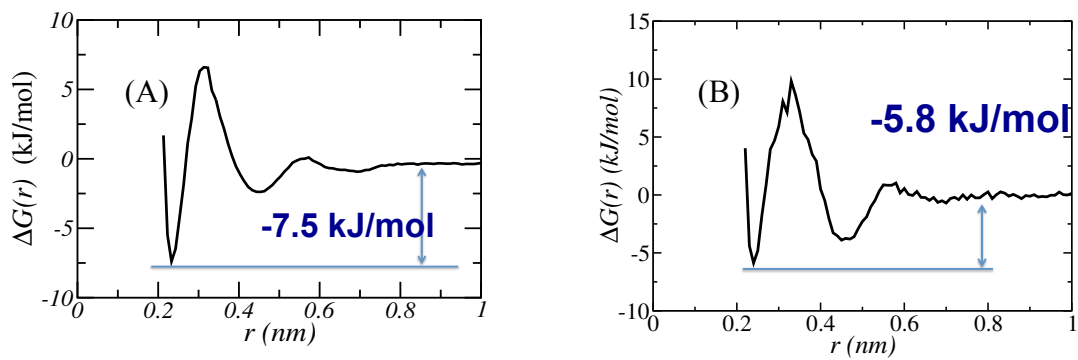

Figure 6. The binding free energy profiles, $\Delta G(r)$, between sulfonate group and (A) $Na^+$ and (B) $Ca^{2+}$ from MD simulations.

The plots show that the general shapes and energy scales of the binding free energy profiles are rather similar between $Na^+$-sulfonate and $Ca^{2+}$-sulfonate ion pairs. The binding strength of the $Na^+$ with sulfonate, defined as the largest depth of the free energy difference relative to the baseline of zero at the farthest distance, is slightly stronger at -7.5 kJ/mol than for the case of $Ca^{2+}$ at -5.8 kJ/mol. The $Ca^{2+}$ binding to the sulfonate has somewhat larger free energy barrier between SIP and CIP, making the transition between two states less favorable. Compared with the $Na^+$-sulfate case, the $Ca^{2+}$-sulfonate case has a slightly deeper minimum in the binding free energy near $r = 0.46$ nm for SIP state, as already mentioned earlier in the discussion for the radial distribution functions. The binding free energy of -7.5 and -5.8 kJ/mol for $Na^+$-sulfonate and $Ca^{2+}$-sulfonate ion pairs, respectively, is relatively weak, similar to or smaller than the typical energy scales associated with water hydrogen bonds.[27] The weak binding of sulfonate-metal ion pairs suggests that the impact of the sulfonate-metal cation ion pairs on the overall system behavior at a macroscopic scale is likely to be small and limited in general. As a result, the main consequence of $Na^+$ and $Ca^{2+}$ metal cations dissolved in sulfonate systems should be mostly the simple screening effect of electrostatic interactions. In the case of $Na^+$ the effect of ion condensation is weak but not entirely ignorable due to the high concentration of $Na^+$ in brine.[10b]

2. Cation-carboxylate ion pairs

The ion pairs of Na$^+$ and Ca$^{2+}$ metal cations with the negatively charged carboxylate group in an acetate anion under brine condition were studied by using regular MD simulation. Figure 7 shows the radial distribution functions of the Na$^+$ and Ca$^{2+}$ around the carboxylate oxygen atoms in the acetate from a regular MD simulation. The results reveal that both Na$^+$ and Ca$^{2+}$ ions bind much stronger to the carboxylate than to sulfonate as manifested by the large peak intensities for the CIPs in figure 7.

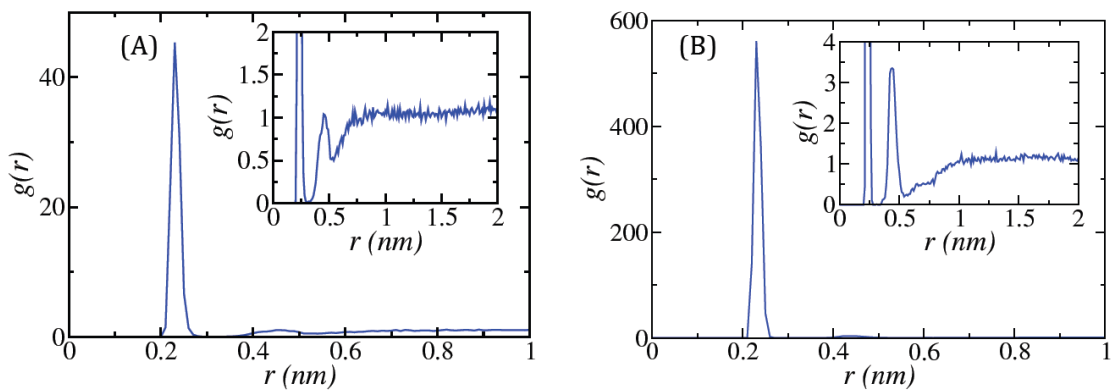

Figure 7. The radial distribution functions of metal ions around acetate for (A) Na$^+$ and (B) Ca$^{2+}$ in the API brine from MD simulation. The insets are the blow-ups for the SIP peaks.

The atomistic simulation provides information on the local structure of the ion pairs and the surroundings. For example, the simulation of an acetate with a mixture of 1.5 M NaCl and 0.18 M CaCl$_2$ reveals a layered structure of Na$^+$ and Ca$^{2+}$ cations and Cl$^-$ anions in an alternating fashion around the carboxylate group (figure 8). Also, there seems to be a local depletion of the metal cations right outside the SIP layers at around $r$ = 0.5 - 0.8 nm, which is compensated by a larger local population of Cl$^-$ ions, possibly to maintain the local overall charge neutrality.

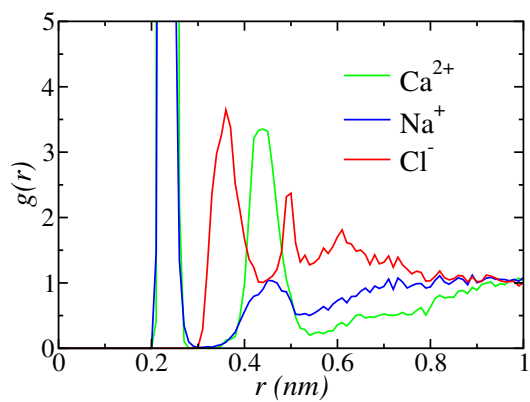

Figure 8. The radial distributions of $Ca^{2+}$, $Na^+$, and $Cl^-$ around carboxylate from MD simulation.

The strong binding of the $Na^+$ and $Ca^{2+}$ to the carboxylate group in the acetate are confirmed by the long correlation times for the CIPs in both $Na^+$ and $Ca^{2+}$ from the residence time correlation functions as shown in figure 9. The correlation functions also show that $Ca^{2+}$ also forms a long-lived SIP with a time scale in the order of 1ns.

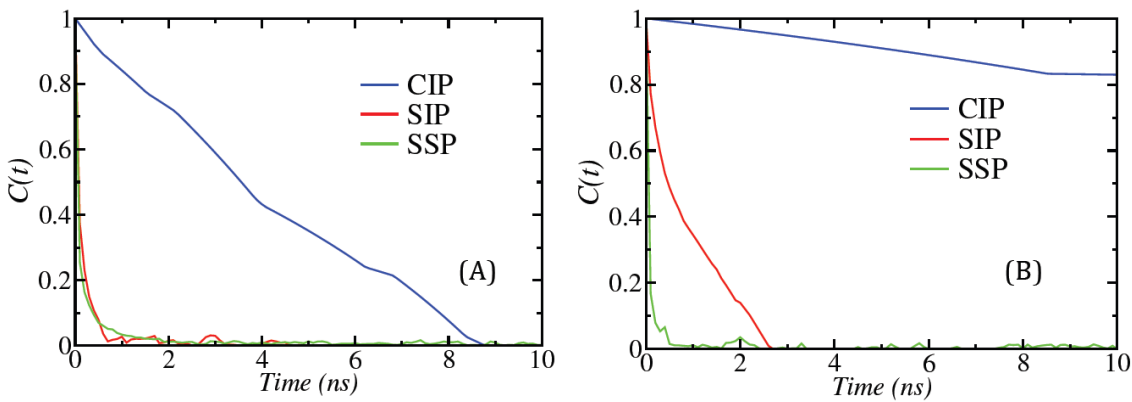

Figure 9. The residence time correlation functions of the ion pairs between acetate and (A) $Na^+$, (B) $Ca^{2+}$ from the regular MD simulation.

As with the sulfonate-metal cation cases, the temporal profiles of the distances between the metal cations and the carboxylate over the simulation time, as shown in figure 10, directly reveal the persistent time scales of the ion pairs in the cation-carboxylate system. Specifically, the time trajectories of the cation-carboxylate distances show that the time scale of the CIP for $Na^+$ with carboxylate is in the order of a few nanoseconds, while for $Ca^{2+}$ in the order of 100 ns or longer, both consistent with the time correlation functions in figure 9.

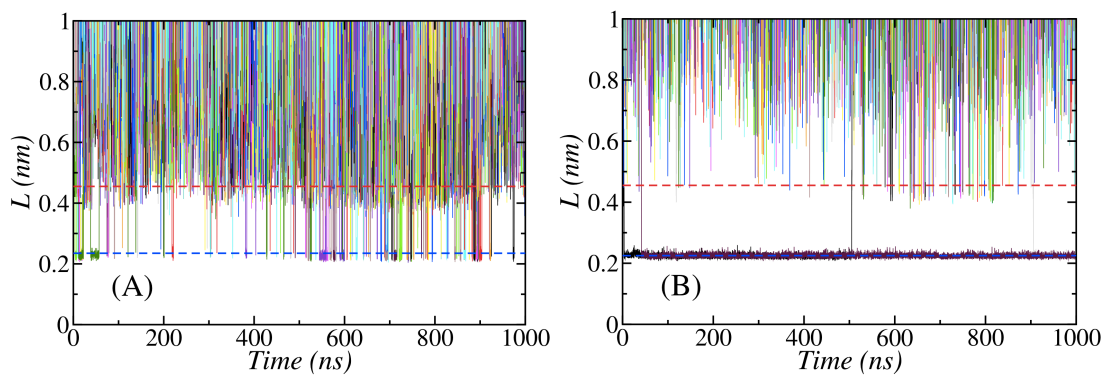

Figure 10. The temporal profiles of the metal-acetate distances for (A) $Na^+$ and (B) $Ca^{2+}$ from MD simulation.

The long correlation time of the ion pairs for $Ca^{2+}$ relative to the total simulation time poses a challenge to the proper sampling of the phase space in regular MD simulation. The sampling quality is critical to the accuracy of the calculation of the binding free energy. To overcome this difficulty and to obtain accurate binding free energies for the metal cation-acetate ion pair systems, we improved the sampling of the ion pairing process for the metal-acetate system by employing constrained MD simulation with umbrella sampling. Detailed procedure of the technique is described in the Simulation Methods section. The technique allows the calculation of the potential of mean force (PMF), which is equivalent to the thermodynamic free energy. The results are shown in figure 11 for the binding between one acetate anion and one metal cation.

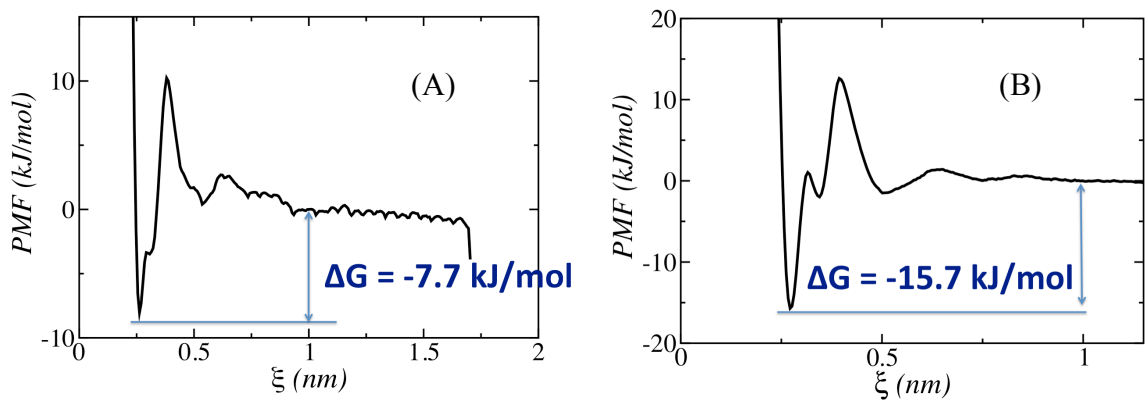

Figure 11. The potential of mean force, $PMF$, and the binding free energy, $\Delta G(r)$, between an acetate anion and (A) $Na^+$ and (B) $Ca^{2+}$ from constrained MD simulations with umbrella sampling.

Interestingly, the binding free energy for $Na^+$-acetate ion pair is calculated to be only -7.7 kJ/mol, which is quite weak and not very different from the sulfonate case. The reason is not clear for such a relatively weak binding free energy for the $Na^+$-acetate ion pair despite the large CIP peak in the radial distribution function as well as the relatively long residence time of almost 5 ns in the correlation function. In the case of the binding between $Ca^{2+}$ and acetate, the association free energy of -15.7 kJ/mol is much stronger than for the sulfonate case. Such a strong binding and the divalent nature of the calcium ion suggest a possible formation of an ionic bridge structure between two negatively charged acetate anions and a single $Ca^{2+}$ cation. We tested the feasibility of such 2:1 binding events by calculating the free energy of ion pairing between a single $Ca^{2+}$ ion and two acetate anions. This was done by first running a series of constrained MD simulations with umbrella sampling of the binding events between a $Ca^{2+}$-acetate 1:1 complex and another acetate anion. The results are shown in figure 12.

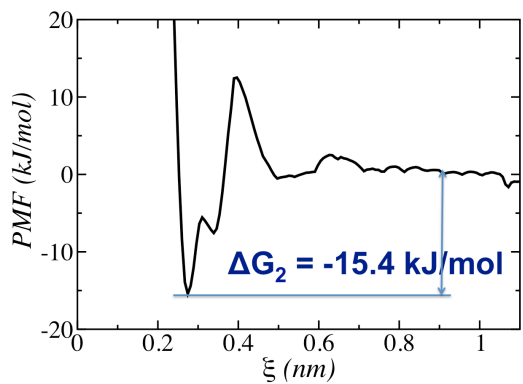

Figure 12. The PMF between a Ca(acetate)$^+$ 1:1 complex and another acetate, obtained from constrained MD simulations with umbrella sampling.

The overall 2:1 binding event of interest ($RCOO^-$ = acetate),

$$Ca^{2+} + 2RCOO^- \rightarrow Ca(RCOO)_2, \Delta G$$

can be rewritten as the sum of two separate, sequential 1:1 binding events as follows.

$$Ca^{2+} + RCOO^- \rightarrow Ca(RCOO)^+, \Delta G_1$$

$$Ca(RCOO)^+ + RCOO^- \rightarrow Ca(RCOO)_2, \Delta G_2$$

Hence the overall free energy, $\Delta G$, associated with the 2:1 binding event between one $Ca^{2+}$ and two acetate anions is $\Delta G = \Delta G_1 + \Delta G_2$, or (-15.7 kJ/mol) + (-15.4 kJ/mol) = -31.1 kJ/mol. The results show a large free energy gain of -31.1 kJ/mol with the 2:1 binding events between a $Ca^{2+}$ and two acetates, so the system should strongly favor the formation of such 2:1 complexes whenever possible.

Finally, we tested the sensitivity of the free energy values for the $Ca^{2+}$-acetate systems to the choice of force fields and water models in the simulations. To do so, we executed multiple MD runs using different combinations among two well-known empirical force fields (AMBER and OPLS-AA) and three popular water models (SPC, TIP3P, and TIP4P) and compared the results among themselves as well as with the experimental values found in the literature. The comparison is compiled in the figure 13. Overall, the calculated binding free energies varied widely among different combinations of force fields and water models, with some calculated values better distributed around the experimental values. In the 1:1 binding case, the parameters from the AMBER force field combined with TIP3P water model produced the best agreement with the experimental values, while the OPLS-AA force field parameters consistently overestimated the binding strength significantly regardless of the choice of the water model. In the case of the 2:1 bindings, the AMBER parameters underestimated the binding strength, while the OPLS-AA parameters again overestimated the interaction significantly compared to the experimental values.

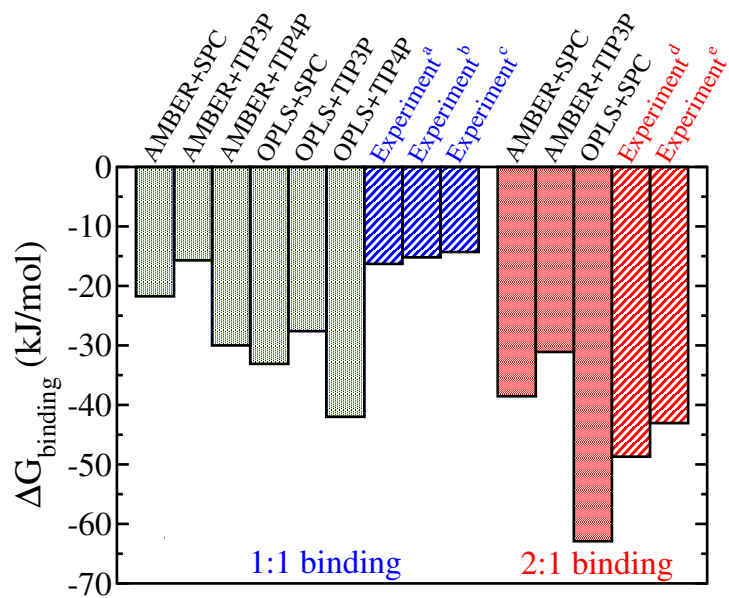

Figure 13. Compilation of the binding free energy values between $Ca^{2+}$ and acetates obtained from the constrained MD simulations with different combinations among two force fields and three water models. Also shown are the experimental values from the literature. (References a, b, c, d, and e in the figure corresponds, respectively, to references [28], [29], [30], [31], and [32].)

Overall the constrained MD simulations with umbrella sampling have provided the binding free energy values that fall roughly within 30% of the experimental values. The comparison highlights both the advantages and the risks of using computational methods in quantifying binding free energy, and for that matter in calculating any physical quantity of interest in general. When a quantity of interest is difficult to obtain experimentally, the computational techniques can be a valuable tool to identify a rough estimate of the value of interest. However the large variability of the computed results in figure 13 is an instructive reminder that the parameters in the popular empirical force fields are sometimes not adequate to produce quantitatively accurate results for a certain physical quantities, and that one should always be vigilant to possible errors due to inadequate and imperfect nature of the parameters in the force fields. In fact, similar observations were reported in the recent studies where the empirical force fields such as GROMOS and OPLS-AA were found to generate very large deviations from experimental values in the PMF calculations for similar ion pairs and some modifications of the standard force field parameters had to be made in order to produce decent results.[33]

When the results from computational approach are compared with the values from experiments, another difficulty arises when the experimental measurements may reflect slightly different conditions than what the computational work represents. Experimental systems are often complex, and microscopic fluctuations, noise, and internal variations sometimes can influence the measurements. Model systems in computational studies, on the other hand, are often very clean and well defined. The differences between what is calculated computationally and what is measured by experiments can sometimes be subtle but significant. In the case of the association free energy in the ion pairs, the experiments for 2:1 binding between $Ca^{2+}$ and acetate may not have considered the rigorous stoichiometry of the association events properly at an atomic and a molecular level. Instead, they may have measured the binding energy based on an average total amount of $Ca^{2+}$ bound to carboxylate in the experiments, which may or may not be the same as the binding energy for molecularly precise 2:1 binding events only. The binding energy measured in experiments is the ensemble and time average of not only the 2:1 complex but many different possible complexes of different assemblies of different number of $Ca^{2+}$, water, acetate and $Cl^-$ ions, similar to the number of different structures shown in figure 3 for $Na^+$ and $Cl^-$ interacting with sulfonate.

CONCLUSIONS

The strong binding between $Ca^{2+}$ and carboxylate is confirmed from our simulations. The multivalent ions are known to dramatically affect the properties of polyelectrolytes, and our results are highly suggestive that similar effects should be in play in the (PAMPS-AA)-coated nanoparticles under the brine condition.

We found a strong preference to form stable bidentate-coordinated contact ion pairs between $Ca^{2+}$ and carboxylate. Similar ion-bridge structures have also been reported in the density functional theory calculations of humic acid[34] and in the MD simulations of natural organic matter.[35] Other multivalent cations such as $Ru(NH_3)_6^{3+}$ were reported to have similar effects of forming ion bridges inside polyelectrolyte brush.[36] In the context of nanoparticles coated with PAMPS-AA, the formation of such a 2:1 ion bridge complex can dramatically change the properties of the nanoparticles, such as dispersion stability and adhesion, by neutralizing the net charge at the local carboxylate groups in the AA monomers as well as by drastically reducing the steric repulsions of the tethered polyelectrolyte layer through the collapse of the polyelectrolyte chains. Therefore, in any theoretical framework to describe and predict the properties of systems that contain $Ca^{2+}$ ions and carboxylate functional group, including both 1:1 and 2:1 binding events between $Ca^{2+}$ and acetates in the theoretical formulation is critically important in order to accurately predict the system properties. In the case of the (PAMPS-AA)-coated nanoparticles in brine in the oil fields, which are the main motivation of this study, the consequences of the strong binding between $Ca^{2+}$ and carboxylate can range from nanoparticle destabilization and aggregations in the bulk brine solution to enhanced or reduced adhesions of nanoparticles to mineral surfaces, leading to a significant change in the nanoparticle transport and retention behavior in the oil reservoir.

ACKNOWLEDGMENT

This research was financially supported by the Advanced Energy Consortium. This research was supported in part through the computational resources and staff contributions provided for the Quest high performance computing facility at Northwestern University which is jointly supported by the Office of the Provost, the Office for Research, and Northwestern University Information Technology.